# RTED: A Robust Algorithm for the Tree Edit Distance

Mateusz Pawlik
Free University of Bozen-Bolzano
Bolzano, Italy
mpawlik@unibz.it

Nikolaus Augsten
Free University of Bozen-Bolzano
Bolzano, Italy
augsten@inf.unibz.it

## ABSTRACT

We consider the classical tree edit distance between ordered labeled trees, which is defined as the minimum-cost sequence of node edit operations that transform one tree into another. The state-of-the-art solutions for the tree edit distance are not satisfactory. The main competitors in the field either have optimal worst-case complexity, but the worst case happens frequently, or they are very efficient for some tree shapes, but degenerate for others. This leads to unpredictable and often infeasible runtimes. There is no obvious way to choose between the algorithms.

In this paper we present RTED, a robust tree edit distance algorithm. The asymptotic complexity of RTED is smaller or equal to the complexity of the best competitors for any input instance, i.e., RTED is both efficient and worst-case optimal. We introduce the class of LRH (Left-Right-Heavy) algorithms, which includes RTED and the fastest tree edit distance algorithms presented in literature. We prove that RTED outperforms all previously proposed LRH algorithms in terms of runtime complexity. In our experiments on synthetic and real world data we empirically evaluate our solution and compare it to the state-of-the-art.

## 1. INTRODUCTION

Tree structured data appears in many applications, for example, XML documents can be represented as ordered labeled trees. An interesting query computes the difference between two trees. This is useful when dealing with different versions of a tree (for example, when synchronizing file directories or archiving Web sites), or to find pairs of similar trees (for example, for record linkage or data mining). The standard approach to tree differences is the tree edit distance, which computes the minimum-cost sequence of node edit operations that transform one tree into another. The tree edit distance has been applied successfully in a wide range of applications, such as bioinformatics [1, 20, 26], image analysis [6], pattern recognition [23], melody recognition [19], natural language processing [25], or information extraction [21, 14], and has received considerable attention from the database community [2, 8, 12, 13, 17, 18, 24].

The tree edit distance problem has a recursive solution that decomposes the trees into smaller subtrees and subforest. The best known algorithms are dynamic programming implementations of this recursive solution. Only two of these algorithms achieve $O(n^2)$ space complexity, where $n$ is the number of tree nodes: The classical algorithm by Zhang and Shasha [31], which runs in $O(n^4)$ time, and the algorithm by Demaine et al. [15], which runs in $O(n^3)$ time. Demaine's algorithm was shown to be worst-case optimal, i.e., no implementation of the recursive solution can improve over the cubic runtime for the worst case. The runtime complexity is given by the number of subproblems that must be solved.

The runtime behavior of the two space-efficient algorithms heavily depends on the tree shapes and it is hard to choose between the algorithms. Zhang's algorithm runs efficiently in $O(n^2 \log^2 n)$ time for trees with depth $O(\log n)$, but it runs into the $O(n^4)$ worst case for some other shapes. Demaine's algorithm is better in the worst case, but unfortunately the $O(n^3)$ worst case happens frequently, also for tree shapes for which Zhang's algorithm is almost quadratic. The runtime complexity can vary by more than a polynomial degree, and the choice of the wrong algorithm leads to a prohibitive runtime for tree pairs that could otherwise be computed efficiently. There is no easy way to predict the runtime behavior of the algorithms for a specific pair of trees, and this problem has not been addressed in literature.

In this paper we develop a new algorithm for the tree edit distance called RTED. Our algorithm is robust, i.e., independent of the tree shape, the number of subproblems that RTED computes is at most as high as the number of subproblems the best competitor must compute. In many cases RTED beats the competitors and is still efficient when they run into the worst case. RTED requires $O(n^2)$ space as the most space-efficient competitors, and its runtime complexity of $O(n^3)$ in the worst case is optimal.

The key to our solution is our dynamic decomposition strategy. A decomposition strategy recursively decomposes the input trees into subforests by removing nodes. At each recursive step, either the leftmost or the rightmost root node must be removed. Different choices lead to different numbers of subproblems and thus to different runtime complexities. Zhang's algorithm always removes from the right, Demaine removes the largest subtree in a subforest last, after removing all nodes to the left and then all nodes to the right of that subtree. Our algorithm dynamically chooses one of the above strategies, and we show that the choice is optimal.

We develop a recursive cost formula for the optimal strategy and we present an algorithm that computes the strategy in $O(n^2)$ time and space. The computation of the strategy does not increase space or runtime complexity of the tree edit distance algorithm. Our experimental evaluation confirms the analytic results and shows that the time used to compute the strategy is small compared to the overall runtime, and the percentage decreases with the tree size. Summarizing, the contribution of this paper is the following:

- We introduce the class of LRH algorithms and the general tree edit distance algorithm, GTED, which implements any LRH strategy in $O(n^2)$ space.

- We present an efficient algorithm that computes the optimal LRH strategy for GTED in $O(n^2)$ time and space. The strategy computation does not increase the overall space or runtime complexity and takes only a small percentage of the overall runtime.

- We present RTED, our robust tree edit distance algorithm. For any tree pair, the number of subproblems computed by RTED is at most the number of subproblems computed by any known LRH algorithm.

- We empirically evaluate RTED and compare it to the state-of-the-art algorithms. To the best of our knowledge, this is the first experimental evaluation of the state-of-the-art in computing the tree edit distance.

The rest of the article is organized as follows. Section 2 provides background material, Section 3 defines the problem, and Section 4 introduces GTED. In Section 5 we analyze decomposition strategies for GTED and introduce the robust tree edit distance algorithm RTED in Section 6. Section 7 discusses related work. We experimentally evaluate our solution in Section 8 and conclude in Section 9.

## 2. NOTATION AND BACKGROUND

We introduce our notation and recap basic concepts of the the tree edit distance computation.

### 2.1 Notation

A *tree* $T$ is a directed, acyclic, connected graph with nodes $N(T)$ and edges $E(T) \subseteq N(T) \times N(T)$, where each node has at most one incoming edge. A *forest* $F$ is a graph in which each connected component is a tree; each tree is also a forest. Each node has a *label*, which is not necessarily unique. The nodes of a forest $F$ are strictly and totally ordered such that $v > w$ for any edge $(v, w) \in E(F)$. The tree traversal that visits all nodes in ascending order is the *postorder* traversal.

In an edge $(v, w)$, node $v$ is the *parent* and $w$ is the *child*, $p(w) = v$. A node with no parent is a *root* node, a node without children is a *leaf*. A node $x$ is an ancestor of node $v$ iff $x = p(v)$ or $x$ is an ancestor of $p(v)$; $x$ is a *descendant* of $v$ iff $v$ is an ancestor of $x$. A node $v$ is *to the left* (right) of $w$ iff $v < w$ ($v > w$) and $v$ is not a descendant (ancestor) of $w$. $r_L(F)$ and $r_R(F)$ are respectively the leftmost and rightmost root nodes in $F$; if $F$ is a tree, then $r(F) = r_L(F) = r_R(F)$.

A *subforest* of a tree $T$ is a graph with nodes $N' \subseteq N(T)$ and edges $E' = \{(v, w) \mid (v, w) \in E(T), v \in N', w \in N'\}$. $T_v$ is the *subtree rooted in node* $v$ of $T$ iff $T_v$ is a subforest of $T$ and $N(T_v) = \{x \mid x = v \text{ or } x \text{ is a descendant of } v \text{ in } T\}$. A *path* $\gamma$ in $F$ is a connected subforest of $F$ in which each node has at most one child.

We use the following short notation: By $|F| = |N(F)|$ we denote the size of $F$, we write $v \in F$ for $v \in N(F)$, and denote the empty forest with $\emptyset$. $F - v$ is the forest obtained from $F$ by removing node $v$ and all edges at $v$. By $F - F_v$ we denote the forest obtained from $F$ by removing subtree $F_v$. $F - \gamma$ is the set of subtrees of $F$ obtained by removing path $\gamma$ from $F$: $F - \gamma = \{F_v : v \notin \gamma \land \exists x(x \in \gamma \land x = p(v))\}$.

EXAMPLE 1. *The nodes of tree $T$ in Figure 1 are $N(T) = \{v_1, v_2, v_3, v_4, v_5\}$, the edges are $E(T) = \{(v_1, v_2), (v_1, v_5), (v_1, v_4), (v_5, v_3)\}$, the node labels are shown in italics in the figure. The root of $T$ is $r(T) = v_1$, and $|T| = 5$. $T_{v_5}$ with nodes $N(T_{v_5}) = \{v_5, v_3\}$ and edges $E(T_{v_5}) = \{(v_5, v_3)\}$ is a subtree of $T$. $T - v_1$ is the subforest of $T$ with $N(T - v_1) = \{v_2, v_3, v_4, v_5\}$ and $E(T - v_1) = \{(v_5, v_3)\}$, $r_L(T - v_1) = v_2$, $r_R(T - v_1) = v_4$.*

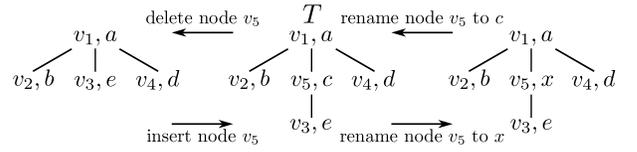

Figure 1: Example trees and edit operations.

### 2.2 Recursive Solution for Tree Edit Distance

The tree edit distance, $\delta(F, G)$, is defined as the minimum-cost sequence of node edit operations that transforms $F$ into $G$. We use the standard edit operations [15, 31]: *delete* a node and connect its children to its parent, maintaining the order; *insert* a new node between an existing node, $v$, and a consecutive subsequence of $v$'s children; *rename* the label of a node (see Figure 1). The costs are $c_d(v)$ for deleting $v$, $c_i(v)$ for inserting $v$, and $c_r(v, w)$ for renaming $v$ to $w$.

The tree edit distance has the recursive solution shown in Figure 2 [31]. The distance between two forests $F$ and $G$ is computed in constant time from the solutions of three or four (depending on whether the two forests are trees) of the following smaller subproblems: (1) $\delta(F - v, G)$, (2) $\delta(F, G - w)$, (3) $\delta(F_v, G_w)$, (4) $\delta(F - F_v, G - G_w)$, and (5) $\delta(F - v, G - w)$. The nodes $v$ and $w$ are either both the leftmost ($v = r_L(F)$, $w = r_L(G)$) or both the rightmost ($v = r_R(F)$, $w = r_R(G)$) root nodes of the respective forests. The subproblems that result from recursively decomposing $F$ and $G$ are called the *relevant subproblems*. The number of relevant subproblems depends on the choice of the nodes $v$ and $w$ at each recursive step.

### 2.3 Dynamic Programming Algorithms

The fastest algorithms for the tree edit distance are dynamic programming implementations of the recursive solution. Since each subproblem is computed in constant time from other subproblems, the runtime complexity of these algorithms is equal to the number of different relevant subproblems they produce. Decomposing a tree with the recursive formula in Figure 2 can result in a quadratic number of subforests. Thus the space complexity of a straight forward algorithm, which stores the distance between all pairs of subforests of two trees $F$ and $G$, is $O(|F|^2|G|^2)$.

Fortunately, the required storage space can be reduced to $O(|F||G|)$ by computing the subproblems bottom-up and reusing space. To achieve this goal, the quadratic space

$$\delta(\emptyset, \emptyset) = 0,$$
$$\delta(F, \emptyset) = \delta(F - v, \emptyset) + c_d(v),$$
$$\delta(\emptyset, G) = \delta(\emptyset, G - w) + c_i(w),$$

if $F$ is not a tree or $G$ is not a tree:
$$\delta(F, G) = \min \begin{cases} \delta(F - v, G) + c_d(v) & (1) \\ \delta(F, G - w) + c_i(w) & (2) \\ \delta(F_v, G_w) + \delta(F - F_v, G - G_w) & (3), (4) \end{cases}$$

if $F$ is a tree and $G$ is a tree:
$$\delta(F, G) = \min \begin{cases} \delta(F - v, G) + c_d(v) & (1) \\ \delta(F, G - w) + c_i(w) & (2) \\ \delta(F - v, G - w) + c_r(v, w) & (5) \end{cases}$$

**Figure 2: Recursive formula for Tree Edit Distance.**

solutions restrict the choice of $v$ and $w$ in each recursive step. Two solutions have been presented in literature. The algorithm by Zhang and Shasha [31] always chooses the same direction ($v$ and $w$ are either both leftmost or both rightmost root nodes). Demaine et al. [15] switch between left- and rightmost root nodes depending on a predefined path.

While the recursive formula is symmetric, i.e., $\delta(F, G)$ and $\delta(G, F)$ produce the same set of subproblems, this does not necessarily hold for its dynamic programming implementations. The bottom-up strategies used in these algorithms compute and store subproblems for later use. When the direction is allowed to change, the subproblems that are required later are hard to predict and only a subset of the precomputed subproblems is actually used later. Thus, in addition to the direction, a decomposition strategy must also choose the order of the parameters.

## 3. PROBLEM DEFINITION

As outlined in the previous section, a dynamic programming algorithm that implements the recursive solution of the tree edit distance must choose a direction (left or right) and the order of the input forests at each recursive step. The choices at each step form a strategy, which determines the overall number of subproblems that must be computed.

In this paper we introduce the class of *path strategies* (cf. Section 4). Path strategies can be expressed using a set of non-overlapping paths that connect tree nodes to leaves. The choice at each recursive step depends on the position of the paths. The class of path strategies is of particular interest since only for this class quadratic space solutions are known. An *LRH strategy* is a path strategy which uses only left, right, and heavy paths. LRH strategies are sufficient to express strategies with optimal asymptotic complexity. An algorithm based on an LRH strategy is an *LRH algorithm*. The most efficient tree edit distance algorithms presented in literature [15, 22, 31] fall into the class of LRH algorithms.

The state-of-the-art in path algorithms is not satisfactory. All previously proposed path algorithms degenerate, i.e., they run into their worst case although a better path strategy exists. The difference can be a polynomial degree, leading to highly varying and often infeasible runtimes.

The goal of this paper is to develop a new LRH algorithm which combines the features of previously proposed algorithms (worst case guarantees for time and space) and in addition is robust, i.e., it does not degenerate. More specifically, the algorithm should have the following properties:

- *space-efficient:* the space complexity should be $O(n^2)$, which is the best known complexity for a tree edit distance algorithm;
- *optimal runtime:* the runtime complexity should be $O(n^3)$, which has shown to be optimal among all possible strategies for the recursive formula in Figure 2 [15];
- *robust:* the algorithm should not run into the worst case if a better LRH strategy exists.

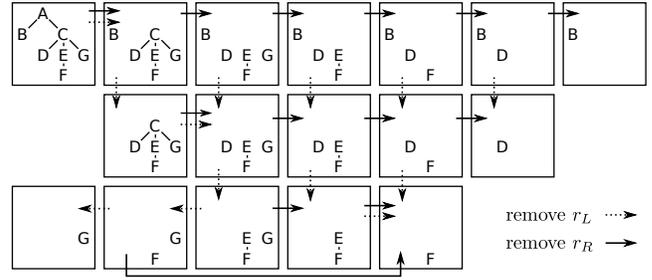

**Figure 3: Subforests of full decomposition.**

Our solution is the RTED algorithm, which satisfies all above requirements. In addition we show that for any instance the number of subproblems computed by RTED is smaller or equal to the number of subproblems computed by all previously proposed LRH algorithms.

## 4. A GENERAL ALGORITHM FOR PATH STRATEGIES

In this section we introduce the GTED algorithm, which generalizes most of the existing tree edit distance algorithms.

### 4.1 Relevant Subforests and Subtrees

The subforests that result from decomposing a tree with the recursive formula in Figure 2 are called the *relevant subforests*. The set of all subforests that can result from any decomposition is called the *full decomposition*. It results from repeated removal of the rightmost or leftmost root node. An example is shown in Figure 3.

*Definition 1.* The *full decomposition* of a tree $F$, $\mathcal{A}(F)$, is the set of all subforests of $F$ obtained by recursively removing the leftmost and rightmost root nodes, $r_L(F)$ and $r_R(F)$, from $F$ and the resulting subforests:

$$\mathcal{A}(\emptyset) = \emptyset$$
$$\mathcal{A}(F) = \{F\} \cup \mathcal{A}(F - r_L(F)) \cup \mathcal{A}(F - r_R(F))$$

Next we show how to decompose a tree into subtrees and subforest based on a so-called *root-leaf path*, a path that connects the root node of the tree to one of its leaves. The set of all root-leaf paths of $F$ is denoted as $\gamma^*(F)$. The left path, $\gamma^L(F)$, the right path, $\gamma^R(F)$, and the heavy path, $\gamma^H(F)$, recursively connect a parent to its leftmost child, its rightmost child, or to the child which roots the largest subtree, respectively. Decomposing trees with root-leaf paths is

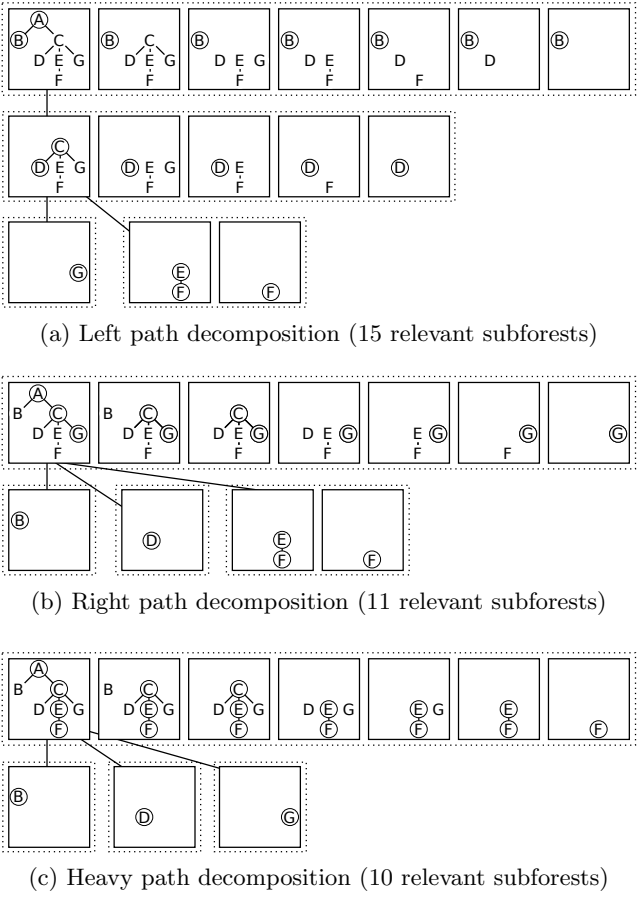

Figure 4: **Relevant subtrees and subforests.**

(a) Left path decomposition (15 relevant subforests)

(b) Right path decomposition (11 relevant subforests)

(c) Heavy path decomposition (10 relevant subforests)

essential for the path strategies discussed in the next subsection.

The *relevant subtrees* of tree $F$ for some root-leaf path $\gamma$ are all subtrees that result from removing $\gamma$ from $F$, i.e., all subtrees of $F$ that are connected to a node on the path $\gamma$.

*Definition 2.* The set of *relevant subtrees* of a tree $F$ with respect to a root-leaf path $\gamma \in \gamma^*(F)$ is defined as $F - \gamma$.

The *relevant subforests* of $F$ for some root-leaf path are $F$ itself and all subforests obtained by removing nodes in the following order: (1) remove the root of $F$ and stop if no nodes are left, (2) remove the leftmost root node in the resulting forest until the leftmost root node is on the root-leaf path, (3) remove the rightmost root node until the rightmost root node is on the root-leaf path, (4) recursively repeat the procedure for the resulting subtree.

*Definition 3.* The set of *relevant subforests* of a tree $F$ with respect to a root-leaf path $\gamma \in \gamma^*(F)$ is recursively defined as

$$\mathcal{F}(\emptyset, \gamma) = \emptyset$$

$$\mathcal{F}(F, \gamma) = \{F\} \cup \begin{cases} \mathcal{F}(F - r_R(F), \gamma) & \text{if } r_L(F) \in \gamma \\ \mathcal{F}(F - r_L(F), \gamma) & \text{otherwise} \end{cases}$$

EXAMPLE 2. *The leftmost tree in each dotted box in Figure 4 is a relevant subtree, and the nodes on the root-leaf path are circled. The other forests in the box are the relevant subforests with respect to the root-leaf path. The solid lines connect trees and their relevant subtrees.*

*Recursive path decomposition.* So far, we have considered relevant subtrees and subforests with respect to a single root-leaf path. If we assume that also for each of the resulting relevant subtrees of a tree $F$ a root-leaf path is defined, we can recursively continue the decomposition as follows: (1) produce all relevant subforest of $F$ with respect to some root-leaf path $\gamma$, (2) recursively apply this procedure to all relevant subtrees of $F$ with respect to $\gamma$.

The set of paths that is used to recursively decompose $F$ is called *path partitioning* of $F$ and is denoted as $\Gamma(F)$. A path partitioning is a set of non-overlapping paths that all end in a leaf node and cover the tree. We define the left path partitioning of $F$ to be $\Gamma^L(F) = \{\gamma^L(F)\} \cup \bigcup_{F_v \in F - \gamma^L(F)} \Gamma^L(F_v)$; the right path partitioning, $\Gamma^R(F)$, is defined analogously.

The set of relevant subforests that result from recursively decomposing tree $F$ with a path partitioning $\Gamma = \Gamma(F)$ is

$$\mathcal{F}(F, \Gamma) = \mathcal{F}(F, \gamma_F) \cup \bigcup_{F' \in F - \gamma_F} \mathcal{F}(F', \Gamma), \quad (1)$$

where $\gamma_F$ is the root-leaf path of $F$ in $\Gamma$, i.e., the only element of $\gamma^*(F) \cap \Gamma$. The set of relevant subtrees that results from recursively decomposing $F$ with the path partitioning $\Gamma$ is $\mathcal{T}(F, \Gamma) = \{F\} \cup \bigcup_{F' \in F - \gamma_F} \mathcal{T}(F', \Gamma)$.

### 4.2 Path Strategies

A *decomposition strategy* chooses between leftmost and rightmost root node at each recursive step, resulting in a set of relevant subforests, which is a subset of the full decomposition. We define a class of decomposition strategies that uses root-leaf paths to decompose trees.

*Definition 4.* A *path strategy* $S$ for two trees $F$ and $G$ maps each pair of subtrees $(F_v, G_w)$, $v \in F$, $w \in G$, to a root-leaf path in one of the subtrees, $\gamma^*(F_v) \cup \gamma^*(G_w)$. An *LRH strategy* is a path strategy that maps subtree pairs to left, right, and/or heavy paths.

Path strategies determine the choice of left vs. right at each step in the recursive tree edit distance solution. A simple algorithm that uses a given path strategy in the recursive solution colors the nodes on the paths and works as follows: Initially, all nodes are white. If both $F$ and $G$ are trees with white root nodes, color the nodes on path $S(F, G)$ (the path to which strategy $S$ maps the pair of subtrees $(F, G)$) in black, all other nodes in $F$ and $G$ are white. Let $B \in \{F, G\}$ be the tree/forest with a colored root node $b$. If the leftmost root of $B$, $r_L(B)$, is black, then $v = r_R(F)$ and $w = r_R(G)$ in the recursive formula, otherwise $v = r_L(F)$ and $w = r_L(G)$.

### 4.3 Quadratic Space Implementation

Space-efficient implementations of the tree edit distance use a bottom-up approach, which computes distances between smaller pairs of subtrees first. By carefully ordering the subtree computations, storage space can be reused without violating the preconditions for later subtree computations. The quadratic space algorithms presented in literature are based on a distance matrix and a quadratic space function which fills the distance matrix.

The *distance matrix* $D$ stores the distances between pairs of subtrees of $F$ and $G$. We represent $D$ as a set of triples

$(F_v, G_w, d)$, where $d = \delta(F_v, G_w)$ is the tree edit distance between $F_v$ and $G_w$, $v \in F$, $w \in G$; the transposed distance matrix is defined as $D^T = \{(G_w, F_v, d) \mid (F_v, G_w, d) \in D\}$.

A *single-path function*, $\Delta(F, G, \gamma_F, D)$, computes distances between the subtrees of two trees $F$ and $G$ for a given root-leaf path $\gamma_F$ in $F$. More specifically, the single-path function between two trees $F$ and $G$ computes the distance between all subtrees of $F$ that are rooted in the root-leaf path $\gamma_F$ and all subtrees of $G$. The precondition is that the distances between all relevant subtrees of $F$ with respect to $\gamma_F$ and all subtrees of $G$ are stored in the distance matrix $D$, $\{(F_v, G_w, d) \mid F_v \in F - \gamma, w \in G, d = \delta(F_v, G_w)\} \subseteq D$, which is part of the input. The output is $D_{out} = D \cup \{(F_v, G_w, d) \mid v \in \gamma_F, w \in G, d = \delta(F_v, G_w)\}$.

Two single-path functions with quadratic space complexity have been proposed in literature. They differ in the types of path (left, right, any) they can process and in the number of subproblems they need to compute. The single-path function by Demaine et al. [15] (called "compute period" in their paper and computed for every node on the path) works for any type of path and computes $|\mathcal{F}(F, \gamma_F)| \times |\mathcal{A}(G)|$ subproblems. The single-path function by Zhang and Shasha [31] (their "tree distance function" can be adapted to be a single-path function) can only process left paths, but needs to compute less subproblems, $|\mathcal{F}(F, \gamma_F)| \times |\mathcal{F}(G, \Gamma^L(G))|$, where $\mathcal{F}(G, \Gamma^L(G)) \subseteq \mathcal{A}(G)$ is the set of subforests obtained by recursively decomposing $G$ using only left root-leaf paths. The symmetric version of this function processes only right paths with similar complexity bounds. We denote the above single-path functions with $\Delta^I$, $\Delta^L$, and $\Delta^R$, respectively.

### 4.4 A General Tree Edit Distance Algorithm

The *general tree edit distance algorithm* (GTED, Algorithm 1) computes the tree edit distance for any path strategy $S$ in quadratic space. The input is a pair of trees, $F$ and $G$, the strategy $S$, and the distance matrix $D$, which initially is empty. The algorithm fills the distance matrix with the distance between all pairs of subtrees from $F$ and $G$ (including $F$ and $G$ themselves).

GTED works as follows. It looks up the root-leaf path $\gamma$ in strategy $S$ for the pair of input trees $(F, G)$. If $\gamma$ is a path in the left-hand tree $F$, then

- GTED is recursively called for tree $G$ and every relevant subtree of tree $F$ with respect to path $\gamma$,
- the single-path function that corresponds to the type of path $\gamma$ is called for the pair of input trees $F$ and $G$.

If $\gamma$ is a path in the right-hand tree $G$, then GTED is called with the trees swapped. Conceptually, this requires to transpose the strategy $S$ and the distance matrix $D$; finally, also the output must be transposed to get the original order of $F$ and $G$ in the distance matrix. In the implementation, the transposition is a flag that indicates the tree order.

The space complexity of GTED is quadratic. Both strategy and distance matrix are of size $|F||G|$. The single-path functions require $O(|F||G|)$ space and release the memory when they terminate. Only one such function is called at the same time. The runtime complexity depends on the strategy and can widely vary between $O(|F||G|)$ and $O(|F|^2|G|^2)$. Below we briefly discuss the strategies of the most important tree edit distance algorithms presented in literature.

### 4.5 Generalization of Previous Work

---

**Algorithm 1:** GTED(F,G,S,D)

1   $\gamma \leftarrow S(F, G)$
2   **if** $\gamma \in \gamma^*(F)$ **then**
3     **forall the** $F' \in F - \gamma$ **do**
4       $D \leftarrow D \cup \text{GTED}(F', G, S, D)$
5     **if** $\gamma = \gamma^L(F)$ **then** $D \leftarrow D \cup \Delta^L(F, G, \gamma, D)$
6     **else if** $\gamma = \gamma^R(F)$ **then** $D \leftarrow D \cup \Delta^R(F, G, \gamma, D)$
7     **else** $D \leftarrow D \cup \Delta^I(F, G, \gamma, D)$
8   **else** $D \leftarrow D \cup (\text{GTED}(G, F, S^T, D^T))^T$
9   **return** $D$

---

Our general tree edit distance algorithm, GTED, generalizes much of the previous work on the tree edit distance, i.e., most algorithms presented in literature are equivalent to GTED with a specific path strategy. We discuss the strategies that turn GTED into the algorithms of Zhang and Shasha [31], Klein [22], and Demaine et al. [15].

The strategy followed by Zhang and Shasha [31] maps all pairs of subtrees $(F_v, G_w)$, $v \in F$, $w \in G$, to the left path, $\gamma^L(F_v)$, resulting in an algorithm with runtime $O(n^4)$ in the worst case. The symmetric strategy maps all pairs of subtrees to the right path, $\gamma^R(F_v)$, with the same runtime complexity. Klein [22] maps all pairs of subtrees to the heavy path, $\gamma^H(F_v)$, achieving runtime $O(n^3 \log n)$.

The strategies discussed above use only the root-leaf paths in $F$ for the decomposition. Demaine et al. [15] use both trees and map all pairs of subtrees $(F_v, G_w)$, $v \in F$, $w \in G$ to $\gamma^H(F_v)$ if $|F_v| \geq |G_w|$ and to $\gamma^H(G_w)$ otherwise. Thus, in each recursive step, the larger tree is decomposed using the heavy path. The resulting algorithm has runtime $O(n^3)$.

## 5. COST OF PATH STRATEGIES

In this section, we count the number of relevant subproblems that must be computed by a single-path function and the overall GTED algorithm. We develop a cost formula, which allows us to count the number of relevant subproblems of the optimal LRH strategy for any pair of input trees.

### 5.1 Relevant Subproblems

A *relevant subproblem* is a pair of relevant subforest, for which a distance must be computed during the recursive tree edit distance evaluation. The number of relevant subproblems depends on the decomposition strategy and determines the runtime complexity of the respective algorithm. The set of relevant subproblems is always a subset of $\mathcal{A}(F) \times \mathcal{A}(G)$.

### 5.2 Complexity of the Single-Path Functions

We count the number of relevant subproblems that the single-path functions $\Delta^L$, $\Delta^R$, and $\Delta^I$ compute for a pair of trees, $F$ and $G$, and a given path $\gamma \in \gamma^*(F)$. We need to derive the number of subforests in the full decomposition of a tree, the decomposition with a single path, and the recursive decomposition with a set of paths.

*Lemma 1.* The number of the subforests in the full path decomposition of tree $F$ is $|\mathcal{A}(F)| = \frac{|F|(|F|+3)}{2} - \sum_{v \in F} |F_v|$.

PROOF. Proof by [16]. □

*Lemma 2.* The number of relevant subforests of tree $F$ with respect to a root-leaf path, $\gamma \in \gamma^*(F)$, is equal to the number of nodes in $F$, $|\mathcal{F}(F,\gamma)| = |F|$.

PROOF. The proof is by induction on the size of $F$. Basis: For $|F| = 1$, $|\mathcal{F}(F,\gamma)| = |F|$ holds due to Definition 3 of $\mathcal{F}(F,\gamma)$. Induction step: We assume, that $|\mathcal{F}(F,\gamma)| = |F|$ holds for all trees $F_k$ of size $k$. Then, for a tree $F_{k+1}$ of size $k+1$ with Definition 3: $|\mathcal{F}(F_{k+1},\gamma)| = |\{F_{k+1}\} \cup \mathcal{F}(F_{k+1} - v, \gamma)| = 1 + |\mathcal{F}(F_{k+1}-v,\gamma)| = 1 + k$ since $|F_{k+1} - v| = k$. □

With the results of Lemma 1 and 2, the cardinalities of both $\mathcal{A}(F)$ and $\mathcal{F}(F,\gamma)$ are independent of the path used to decompose tree $F$ and can be computed in linear time in a single traversal of $F$ for the tree and all its subtrees.

*Lemma 3.* The number of relevant subforests produced by a recursive path decomposition of tree $F$ with path partitioning $\Gamma$ is the sum of the sizes of all the relevant subtrees in the recursive decomposition:

$$|\mathcal{F}(F,\Gamma)| = \sum_{F' \in \mathcal{T}(F,\Gamma)} |F'|$$

PROOF. Follows from the definition of $\mathcal{F}(F,\Gamma)$ in Equation (1) and Lemma 2. □

EXAMPLE 3. *Figure 4 shows different recursive path decompositions. All path decompositions in the figure produce different numbers of relevant subforests, and they all produce less subforests than the full decomposition in Figure 3.*

The next lemma counts the number of relevant subproblems that the different single-path functions must compute.

*Lemma 4.* The number of relevant subproblems computed by the single-path functions $\Delta^I$, $\Delta^L$, and $\Delta^R$, for a pair of trees $F$ and $G$ is as follows ($D$ is the distance matrix):

- $\Delta^I(F,G,\gamma,D)$, $\gamma \in \gamma^*(F)$: $|F| \times |\mathcal{A}(G)|$
- $\Delta^L(F,G,\gamma^L(F),D)$: $|F| \times |\mathcal{F}(G,\Gamma^L(G))|$
- $\Delta^R(F,G,\gamma^R(F),D)$: $|F| \times |\mathcal{F}(G,\Gamma^R(G))|$

PROOF. Demaine et al. [15] show that the number of subproblems computed by $\Delta^I(F,G,\gamma,D)$, $\gamma \in \gamma^*(F)$, is $|\mathcal{F}(F,\gamma)| \times |\mathcal{A}(G)|$. Zhang and Shasha [31] show the number of relevant subproblems for $\Delta^L(F,G,\gamma^L(F),D)$ to be $|\mathcal{F}(F,\gamma^L(F))| \times |\mathcal{F}(G,\Gamma^L(G))|$; $|\mathcal{F}(F,\gamma^L(F))| = |F|$ follows from Lemma 2. The same rationale holds for $\Delta^R$. □

## 5.3 Cost of the Optimal Strategy

With the results in Section 5.2 we can compute the cost of GTED for any path strategy, i.e., the number of relevant subproblems that GTED must compute. The overall cost of a strategy is computed by decomposing a tree into its relevant subtrees according to the paths in the strategy and summing up the costs for executing the single-path function for each pair of relevant subtrees.

We compute the cost of the *optimal* LRH strategy. This is achieved by an exhaustive search in the space of all possible LRH strategies for a given pair of trees. At each recursive step, there are six choices. Either $F$ is decomposed by one of left, right, or heavy path, or $G$ is decomposed. For each of the six choices, the cost of the relevant subtrees that result from the decomposition must be explored. The optimal strategy is computed by the formula in Figure 5, which chooses the minimum cost at each recursive step. The cost formula counts the exact number of relevant subproblems for the optimal LRH strategy between two tree.

$$\text{cost}(F,G) = \\ = \min \begin{cases} |F| \times |\mathcal{A}(G)| + \sum_{F' \in F - \gamma^H(F)} \text{cost}(F',G) \\ |G| \times |\mathcal{A}(F)| + \sum_{G' \in G - \gamma^H(G)} \text{cost}(G',F) \\ |F| \times |\mathcal{F}(G,\Gamma^L(G))| + \sum_{F' \in F - \gamma^L(F)} \text{cost}(F',G) \\ |G| \times |\mathcal{F}(F,\Gamma^L(F))| + \sum_{G' \in G - \gamma^L(G)} \text{cost}(G',F) \\ |F| \times |\mathcal{F}(G,\Gamma^R(G))| + \sum_{F' \in F - \gamma^R(F)} \text{cost}(F',G) \\ |G| \times |\mathcal{F}(F,\Gamma^R(F))| + \sum_{G' \in G - \gamma^R(G)} \text{cost}(G',F) \end{cases}$$

**Figure 5: The cost formula.**

*Theorem 1.* The cost formula in Figure 5 computes the cost of the optimal LRH strategy for the general tree edit distance algorithm GTED.

PROOF. Proof by induction over the structure of $F$ and $G$. In the base case, both $F$ and $G$ consist of a single node and $\text{cost}(F,G) = 1$. Inductive hypothesis: $\text{cost}(F',G)$ and $\text{cost}(F,G')$ for the relevant subtrees $F'$ and $G'$ with respect to the left, right, and heavy path of $F$ and $G$ are optimal. We show that the optimality of $\text{cost}(F,G)$ follows. There are six possible paths for decomposing $F$ and $G$: left, right, or heavy in either $F$ or $G$. If $\gamma$ is a path in $F$, the distance of $G$ to all relevant subtrees of $F$ with respect to $\gamma$ must be computed, which can be done in $\sum_{F' \in F - \gamma} \text{cost}(F',G)$ steps (inductive hypothesis). Further, depending on the path type of $\gamma$, a single path function must be called: $\Delta^L$ for the left path, $\Delta^R$ for the right path, and $\Delta^I$ for the heavy path. The costs follow from Lemma 4 and are added to the cost sum for the relevant subtrees. Calling $\Delta^I$ for left or right paths cannot lead to smaller cost since $\mathcal{F}(F, \Gamma^{L/R}) \subseteq \mathcal{A}(F)$. Similar rationale holds if $\gamma$ is a path in $G$. The cost formula in Figure 5 takes the minimum of the six possible costs. □

The single-path function, $\Delta^I$, which is used for heavy paths in GTED, is not optimal since it computes subproblems that are not required. It is, however, the only known algorithm for heavy paths that runs in $O(n^2)$ space. $\Delta^I$ can easily be substituted with another single-path function by changing the respective costs in the cost formula.

## 6. RTED: ROBUST TREE EDIT DISTANCE ALGORITHM

The *robust tree edit distance algorithm*, RTED, computes the optimal LRH strategy for two trees and runs the GTED algorithm presented in Section 4 with the optimal strategy.

The optimal strategy for two trees, $F$ and $G$, is computed with the cost formula in Figure 5 by memorizing at each recursive step the best root-leaf path. Since there is one path for each pair of relevant subtrees, the paths can be stored in an array of size $|F| \times |G|$, the *strategy array*, which we initialize with empty paths. The strategy array maps each pair of subtrees to a root-leaf path in one of the subtrees, thus the strategy array is a path strategy (cf. Definition 4).

The key challenge is to compute the optimal strategy efficiently. The exhaustive exploration of the search space of exponential size is obviously prohibitive. We observe, however, that the cost is computed only between pairs of subtrees of $F$ and $G$, and there is only a quadratic number of such subtrees. This suggests a dynamic programming approach, which stores the intermediate results and traverses identical branches of the search tree only once.

## 6.1 Baseline Algorithm for Optimal Strategy

The baseline algorithm for computing the optimal strategy implements the cost formula and traces back the optimal strategy. It stores intermediate results and uses dynamic programming to avoid computing the strategy for identical pairs of subtrees more than once. The intermediate results are stored in a *memoization matrix* of size $|F| \times |G|$, and the optimal strategy is stored in the strategy array. At each recursive step, four actions are performed for $F$ and $G$:

- If the cost for $F$ and $G$ is in the memoization matrix, return the cost and ignore the following steps.
- Compute the costs for $G$ ($F$) and the relevant subtrees of $F$ ($G$) w.r.t. the left, right, and heavy paths; sum up the values and compute a cost for each path.
- Store the path with smallest cost as the entry for $F$ and $G$ in the strategy array.
- Store the cost for $F$ and $G$ in the memoization matrix.

*Theorem 2.* The runtime complexity of the baseline algorithm for two trees $F$ and $G$ is bound by $O(n^3)$, $n = \max(|F|, |G|)$, and the bound is tight.

PROOF. The runtime of the baseline algorithm is determined by the number of sums that must be computed. We proceed in three steps: (1) count the number of the summations, (2) show the $O(n^3)$ upper bound for the complexity, (3) give an instance for which this bound is tight.

(1) Summation count: The cost for a subtree pair $(F_v, G_w)$ is the minimum of six values (cf. Figure 5). Computing each value requires $|F - \gamma|$ summations: The cost of the single-path function (product) is computed in constant time since the factors can be precomputed in $O(|F| + |G|)$ time with Lemmas 1 and 3; adding the costs for the relevant subtrees w.r.t. path $\gamma$ requires $|F-\gamma|-1$ summations. Since we store the results, the cost for each pair $(F_v, G_w)$, $v \in F$, $w \in G$, must be computed at most once. For some subtree pairs no cost might be computed, for example, when the root node of one of the subtrees is the only child of its parent. Overall, this leads to an upper bound for the number of summations:

$$\#sums \leq \sum_{v \in F, w \in G} (|F_v - \gamma^L(F_v)| + |F_v - \gamma^R(F_v)| + $$
$$|F_v - \gamma^H(F_v)| + |G_w - \gamma^L(G_w)| + \quad (2)$$
$$|G_w - \gamma^R(G_w)| + |G_w - \gamma^H(G_w)|)$$

(2) Upper bound: The number of relevant subtrees of $F_v$ with respect to $\gamma \in \gamma^*(F_v)$ is limited by the number of leaf nodes of $F_v$, $|F_v - \gamma| \leq |l(F_v)| - 1$. Substituting in (2) we get a new (looser) upper bound, $\#sum \leq \sum_{v \in F, w \in G} (3(|l(F_v)| - 1) + 3(|l(G_w)| - 1))$. There are at most $|F||G|$ different pairs of subtrees, $|l(F_v)| \leq |F|$, $|l(G_w)| \leq |G|$, thus $\#sums \leq |F||G|(3|F| + 3|G|)) = O(n^3)$.

(3) Tightness of bound: We show that for some instances the runtime of the baseline algorithm is $\Omega(n^3)$. Let $F$ be a left branch tree (Figure 7(a)) and $G$ a right branch tree (Figure 7(b)). For a subtree, $F_v$, rooted in a non-leaf $v$ of $F$, it holds that $|F_v - \gamma^L(F_v)| = \frac{|F_v|-1}{2}$, $|F_v - \gamma^H(F_v)| = \frac{|F_v|-1}{2}$, $|F_v - \gamma^R(F_v)| = 1$; if $v$ is a leaf, then $|F_v - \gamma| = 0$ for any path. Similar, for subtree $G_w$ and non-leaf $w$, $|G_w - \gamma^L(G_w)| = 1$, $|G_w - \gamma^H(G_w)| = \frac{|G_w|-1}{2}$, $|G_w - \gamma^R(G_w)| = \frac{|G_w|-1}{2}$; $|G_w - \gamma| = 0$ if $w$ is a leaf. For $F$ and $G$, every pair of subtrees occurs during the computation of the baseline algorithm, thus the right-hand term in (2) is the exact number of summations. By substituting in (2) we get:

$$\#sums = \sum_{v \in F \setminus l(F), w \in G \setminus l(G)} \left(\frac{|F_v|-1}{2} + 1 + \frac{|F_v|-1}{2} + \right.$$
$$\left. 1 + \frac{|G_w|-1}{2} + \frac{|G_w|-1}{2}\right) +$$
$$\sum_{v \in l(F), w \in G \setminus l(G)} \left(1 + \frac{|G_w|-1}{2} + \frac{|G_w|-1}{2}\right) +$$
$$\sum_{v \in F \setminus l(F), w \in l(G)} \left(\frac{|F_v|-1}{2} + 1 + \frac{|F_v|-1}{2}\right)$$

$l(F)$ and $l(G)$ are leaf nodes of $F$ and $G$, respectively. With $|l(F)| = (|F|+1)/2$, $|l(G)| = (|G|+1)/2$, we get $\#sum = \Omega(\frac{|F|}{2}\frac{|G|}{2}(|F|+|G|) + \frac{|F|}{2}\frac{|G|}{2}|G| + \frac{|F|}{2}\frac{|G|}{2}|F|) = \Omega(n^3)$. □

## 6.2 Efficient Algorithm for Optimal Strategy

The runtime of the baseline algorithm is $O(n^3)$. While this is clearly a major improvement over the naive exponential solution, it is unfortunately not enough in our application. The complexity for computing the strategy must not be higher than the complexity of the optimal strategy for GTED. The optimal GTED strategy is often better than cubic, e.g., $O(n^2 \log^2 n)$ for trees of depth $\log(n)$.

In this section we introduce OptStrategy (Algorithm 2), which computes the optimal LRH strategy for GTED in $O(n^2)$ time. Similar to the base line algorithm, a strategy array, $STR$, of quadratic size is used to store the best path at each recursive step, resulting in the best overall strategy.

Different from the base line algorithm, we do not store costs between individual pairs of relevant subtrees. Instead we maintain and incrementally update the cost sums of the relevant subtrees (summations over the relevant subtrees in the cost formula). We do not sum up the same cost multiple times and thus reduce the runtime. The cost sums are stored in *cost arrays*: $L_v$, $R_v$, $H_v$ of size $|F| \times |G|$ store a cost sum for each pair $(F_v, G_w)$, $v \in F$, $w \in G$, for the left, right, and heavy path in $F_v$, respectively; for example, $L_v[F_v, G_w] = \sum_{F' \in F_v - \gamma^L(F_v)} \text{cost}(F', G_w)$. $L_w$, $R_w$, $H_w$ of size $|G|$ store the cost sums between all relevant subtrees of $G$ w.r.t. some path and a specific subtree $F_v$, for example, $L_w[G_w] = \sum_{G' \in G_w - \gamma^L(G_w)} \text{cost}(G', F_v)$.

| Algorithm 2: OptStrategy($F,G$) |
|---|
| 1  $L_v, R_v, H_v$ : arrays of size $|F| \times |G|$ |
| 2  $L_w, R_w, H_w$ : arrays of size $|G|$ |
| 3  **for** $v = 1$ *to* $|F|$ *in postorder* **do** |
| 4    **for** $w = 1$ *to* $|G|$ *in postorder* **do** |
| 5      **if** $w$ *is leaf* **then** $L_w[w] \leftarrow R_w[w] \leftarrow H_w[w] \leftarrow 0$ |
| 6      **if** $v$ *is leaf* **then** $L_v[v,w] \leftarrow R_v[v,w] \leftarrow H_v[v,w] \leftarrow 0$ |
| 7      $C \leftarrow \{(|F_v| \times |\mathcal{A}(G_w)| + H_v[v,w], \gamma^H(F)),$ |
| 8          $(|G_w| \times |\mathcal{A}(F_v)| + H_w[w], \gamma^H(G)),$ |
| 9          $(|F_v| \times |\mathcal{F}(G_w, \Gamma_L(G))| + L_v[v,w], \gamma^L(F)),$ |
| 10        $(|G_w| \times |\mathcal{F}(F_v, \Gamma_L(F))| + L_w[w], \gamma^L(G)),$ |
| 11        $(|F_v| \times |\mathcal{F}(G_w, \Gamma_R(G))| + R_v[v,w], \gamma^R(F)),$ |
| 12        $(|G_w| \times |\mathcal{F}(F_v, \Gamma_R(F))| + R_w[w], \gamma^R(G))\}$ |
| 13      $(c_{min}, \gamma_{min}) \leftarrow (c, \gamma)$ such that $(c, \gamma) \in C$ and $c = \min\{c' \mid (c', \gamma') \in C\}$ |
| 14      $STR[v, w] \leftarrow \gamma_{min}$ |
| 15      **if** $v$ *is not root* **then** |
| 16        $L_v[p(v), w] \stackrel{+}{=} \begin{cases} L_v[v,w] & \text{if } v \in \gamma^L(F_{p(v)}) \\ c_{min} & \text{otherwise} \end{cases}$ |
| 17        $R_v[p(v), w] \stackrel{+}{=} \begin{cases} R_v[v,w] & \text{if } v \in \gamma^R(F_{p(v)}) \\ c_{min} & \text{otherwise} \end{cases}$ |
| 18        $H_v[p(v), w] \stackrel{+}{=} \begin{cases} H_v[v,w] & \text{if } v \in \gamma^H(F_{p(v)}) \\ c_{min} & \text{otherwise} \end{cases}$ |
| 19      **if** $w$ *is not root* **then** |
| 20        $L_w[p(w)] \stackrel{+}{=} \begin{cases} L_w[w] & \text{if } w \in \gamma^L(G_{p(w)}) \\ c_{min} & \text{otherwise} \end{cases}$ |
| 21        $R_w[p(w)] \stackrel{+}{=} \begin{cases} R_w[w] & \text{if } w \in \gamma^R(G_{p(w)}) \\ c_{min} & \text{otherwise} \end{cases}$ |
| 22        $H_w[p(w)] \stackrel{+}{=} \begin{cases} H_w[w] & \text{if } w \in \gamma^H(G_{p(w)}) \\ c_{min} & \text{otherwise} \end{cases}$ |
| 23  **return** $STR$ |

The algorithm loops over every pair of subtrees $(F_v, G_w)$ in postorder of the nodes $v \in F$, $w \in G$. The cost sum for a leaf node is zero, because leaves have no relevant subtrees (Lines 5–6). The values in the cost arrays are used to compute the costs of the pair $(F_v, G_w)$ for each of the six possible paths in the cost formula (Lines 7–12). For each result a $(cost, path)$ pair is stored in the temporary set $C$. The minimum cost in $C$ is assigned to $c_{min}$, the respective path $\gamma_{min}$ is stored in the strategy array. Finally, the cost sums for the subtree pairs $(F_{p(v)}, G_w)$ and $(F_v, G_{p(w)})$ are updated, where $p(v)$ and $p(w)$ are the parents of $v$ and $w$, respectively. The update value depends on whether $v$ and $w$ belong to the same path as their parent.

EXAMPLE 4. *We use Algorithm 2 to compute the optimal strategy for two trees $F$ and $G$, $N(F) = \{1, 2, 3\}$, $E(F) = \{(3, 1), (3, 2)\}$, $N(G) = \{1, 2\}$, $E(G) = \{(2, 1)\}$; for simplicity, node IDs and labels are identical and correspond to the postorder position in the tree. Figure 6 shows the cost arrays and the strategy array before the last node pair, $v = 3$, $w = 2$, is processed. Array rows and columns are labeled with node IDs, e.g., the cost sum for* the subtree pair $(F_3, G_1)$ *w.r.t. the heavy path in $F_3$ is $H_v[3, 1] = 1$. We compute the missing value in the strategy array STR: Neither $v$ nor $w$ are leaves; with $|G_w| = |\mathcal{A}(G_w)| = |\mathcal{F}(G_w, \Gamma_L(G))| = |\mathcal{F}(G_w, \Gamma_R(G))| = 2$, $|F_v| = 3$, $|\mathcal{A}(F_v)| = |\mathcal{F}(F_v, \Gamma_L(F))| = |\mathcal{F}(F_v, \Gamma_R(F))| = 4$ we compute $C = \{(3 * 2 + 2, \gamma^H(F_3)), (2 * 4 + 0, \gamma^H(G_2)), (3 * 2 + 2, \gamma^L(F_3)), (2 * 4 + 0, \gamma^L(G_2)), (3 * 2 + 2, \gamma^R(F_3)), (2*4+0, \gamma^R(G_2))\} = \{(8, \gamma^H(F_3)), (8, \gamma^H(G_2)), (8, \gamma^L(F_3)), (8, \gamma^L(G_2)), (8, \gamma^R(F_3)), (8, \gamma^R(G_2))\}$. In the example, all costs are identical, and we arbitrarily pick $(c_{min}, \gamma_{min}) = (8, \gamma^H(F_3))$ as the minimum; the missing value in the strategy array is $\gamma^H(F_3)$. Since both $v$ and $w$ are roots, the algorithm terminates and returns the optimal strategy.*

| $v \downarrow$ | $H_v$ | | $L_v$ | | $R_v$ | | STR | |
|---|---|---|---|---|---|---|---|---|
| 1 | 0 | 0 | 0 | 0 | 0 | 0 | $\gamma^H(F_1)$ | $\gamma^H(F_1)$ |
| 2 | 0 | 0 | 0 | 0 | 0 | 0 | $\gamma^H(F_2)$ | $\gamma^H(F_2)$ |
| 3 | 1 | 2 | 1 | 2 | 1 | 2 | $\gamma^H(F_3)$ | |
| $w \rightarrow$ | 1 | 2 | 1 | 2 | 1 | 2 | 1 | 2 |

| | $H_w$ | | $L_w$ | | $R_w$ | |
|---|---|---|---|---|---|---|
| | 0 | 0 | 0 | 0 | 0 | 0 |
| $w \rightarrow$ | 1 | 2 | 1 | 2 | 1 | 2 |

$$\begin{array}{cc} 3 & 2 \\ 1 \ 2 & 1 \\ F & G \end{array}$$

**Figure 6: Cost arrays and strategy array.**

*Theorem 3.* Algorithm 2 is correct.

PROOF. The strategy array *STR* maps every pair of subtrees to a root-leaf path and thus is a strategy according to Definition 4. Next we show by induction that the cost arrays store the correct values according to the cost formula. *Base case*: For pairs of leaf nodes $(v, w)$ the cost arrays store zeros; this is correct due to $F_v - \gamma_{F_v} = G_w - \gamma_{G_w} = \emptyset$. *Inductive hypothesis*: For all pairs of children $(v, w)$ of two nodes, $f = p(v)$ and $g = p(w)$, the values in the cost arrays are correct. We show that, after processing all children, the values for $f$ and $g$ are correct. This implies the correctness of the overall algorithm since the nodes are processed in postorder, i.e., all children are processed before their parent.

We consider two cases for node $v$ (for $w$ analogous reasonings hold): (1) node $v$ lies on the same root-leaf path as its parent, (2) node $v$ does not lie on the same root-leaf path as its parent. *Case 1*: $F_v$ is not a relevant subtree of $F_f$ with respect to $\gamma_{F_f}$. The cost already stored for $v$ is the sum of the costs for every relevant subtree $F' \in F_v - \gamma_{F_v}$, i.e., a part of the sum $\sum_{F' \in F_f - \gamma_{F_f}} \text{cost}(F', G_w)$. We increment the value in the cost array of $\gamma_{F_f}$ with the cost already stored for $v$. *Case 2*: $F_v \in F_f - \gamma_{F_f}$ is the root of some relevant subtree of $F_f$ with respect to $\gamma_{F_f}$. The cost of $v$ is an element in the cost sum for $f$, $\sum_{F' \in F_f - \gamma_{F_f}} \text{cost}(F', G_w)$. We add the cost value of $v$ ($c_{min}$) to the cost entry of $f$. □

*Theorem 4.* Time and space complexity of Algorithm 2 are $O(n^2)$, where $n = \max(|F|, |G|)$.

PROOF. Algorithm 2 iterates over all pairs of subtrees $(F_v, G_w)$, $v \in F, w \in G$, thus the innermost loop is executed $|F||G|$ times. In the inner loop we do a constant number of array lookups and sums. The factors of the six products in Lines 7–12 are precomputed in $O(|F| + |G|)$ time and space using the Lemmas in Section 5.2. We use four arrays of size $|F| \times |G|$ and three arrays of size $|G|$, thus the overall complexity is $O(|F||G|)$ in time and space. □

## 7. RELATED WORK

The first tree edit distance algorithm for the unrestricted edit model, which is also assumed in our work, was already proposed in 1979 by Tai [28]. Tai's algorithm runs in $O(m^3n^3)$ time and space for two trees $F$ and $G$ with $m$ and $n$ nodes, respectively. Zhang and Shasha [31] improve the complexity to $O(m^2n^2)$ time and $O(mn)$ space; for trees with $l$ leaves and depth $d$ the runtime is $O(mn \min(l_F, d_F) \min(l_G, d_G))$, which is much better than $O(m^2n^2)$ for some tree shapes. Klein [22] uses heavy paths [27] to decompose the larger tree and gets an $O(n^2m \log m)$ time and space algorithm, $m \geq n$. Demaine et al. [15] also use heavy paths, but different from Klein they switch the trees such that the larger subtree is decomposed in each recursive step. Their algorithm runs in $O(n^2m(1 + \log \frac{m}{n}))$ time and in $O(mn)$ space, $m \geq n$. Although Demaine's algorithm is worst-case optimal [15], it is slower than Zhang's algorithm for some interesting tree shapes, for example, balanced trees. Our RTED algorithm is efficient for all tree shapes for which any of the above algorithms is efficient. The runtime of RTED is worst-case optimal and the space complexity is $O(mn)$.

Similar to our approach, Dulucq and Touzet [16] compute a decomposition strategy in the first step, then use the strategy to compute the tree edit distance. They only consider strategies that decompose a single tree and get an algorithm that runs in $O(n^2m \log m)$ time and space. Our algorithm requires only $O(mn)$ space. Further, we consider strategies that decompose both trees, which adds complexity to the model, but is required to achieve worst-case optimality: our runtime is $O(n^2m(1 + \log \frac{m}{n}))$, which is $O(n^3)$ if $n = m$.

None of the above works includes an empirical evaluation of the algorithms. We evaluate our RTED algorithm on both synthetic and real world data and compare it to the solutions of Zhang and Shasha [31], Klein [22], and Demaine et al. [15]. We further provide a formal framework, which extends previous work by Dulucq and Touzet [16], and an algorithm that generalizes all above approaches.

For specific tree shapes or a restricted set of edit operations faster algorithms have been proposed. Chen [11] presents an $O(mn + l_F^2 n + l_F^{2.5} l_G) = O(m^{2.5}n)$ algorithm based on fast matrix multiplication, which is efficient for some instances, e.g., when one of the trees has few leaves. Chen and Zhang [10] present an efficient algorithm for trees with long chains. By contracting the chains they reduce the size of tree $F$ ($G$) from $m$ to $\tilde{m}$ ($n$ to $\tilde{n}$) and achieve runtime $O(mn + \tilde{m}^2\tilde{n}^2)$. Chawathe [8] restricts the delete operation to leaf nodes. An external memory algorithm that reduces the tree edit distance problem to a well-studied shortest path problem is proposed. Other variants of the tree edit distance are discussed in a survey by Bille [7]. Our algorithm adapts to any tree shape and we assume the unrestricted edit model.

Tree edit distance variants are also used for change detection in hierarchical data. Lee et al. [24] and Chawathe et al. [9] match tree nodes and compute a distance in $O(ne)$ time, where $e$ is the distance between the trees. Cobéna et al. [12] take advantage of element IDs in XML documents, which cannot be generally assumed. The X-Diff algorithm by Wang et al. [29] allows leaf and subtree insertion and deletion, and node renaming. In order to achieve $O(n^2 \times f \log f)$ time complexity for trees with $n$ nodes and maximum fanout $f$, only nodes with the same path to the root are matched.

Lower and upper bounds of the tree edit distance have been studied. Guha et al. [18] propose a lower bound based on the string edit distance between serialized trees and an upper bound based on a restricted tree edit distance variant. Yang et al. [30] decompose trees into so-called binary branches and derive a lower bound from the number of non-matching binary branches between two trees. Similarly, Augsten et al. [4, 5] decompose the trees into pq-grams and provide a lower bound for an edit distance that gives higher weight to nodes with many children. The bounds have more efficient algorithms than the exact tree edit distance that we compute in this paper. Bounds are useful to prune exact distance computations when trees are matched with a similarity threshold. There is no straightforward way to build bounds into the dynamic programming algorithms for the exact tree edit distance to improve its performance.

The TASM (top-$k$ approximate subtree matching) algorithm by Augsten et al. [2] identifies the top-$k$ subtrees in a data tree with the smallest edit distances from a given query tree. TASM prunes distance computations for large subtrees of the data tree and achieves a space complexity that is independent of the data size. The pruning makes use of the top-$k$ guarantee, which is not given in our scenario.

Garofalakis and Kumar [17] embed the tree edit distance with subtree move as an additional edit operation into a numeric vector space equipped with the standard $L_1$ distance norm. They compute an efficient approximation of the tree edit distance with asymptotic approximation guarantees. In our work, we compute the exact tree edit distance.

In this work we assume ordered trees. For unordered trees the problem is NP-hard [32]. Augsten et al. [3] study an efficient approximation for the unordered tree edit distance.

## 8. EXPERIMENTS

We empirically evaluate our solution (RTED) on both real world and synthetic datasets and compare it to the fastest algorithms proposed in literature: the algorithm by Zhang and Shasha [31] (Zhang-L), which uses only left paths to decompose the trees; the symmetric version of this algorithm that always uses right paths (Zhang-R); Klein's algorithm [22] (Klein-H), which uses heavy paths only in one tree; the worst-case optimal solution by Demaine et al. [15] (Demaine-H), which decomposes both trees with heavy paths. Our implementation of Klein's algorithm includes the improvements proposed by Demaine et al. [15] and runs in quadratic space. All algorithms were implemented as single-thread applications in Java 1.6 and run on a dual-core AMD64 server.

*The Datasets.* We test the algorithms on both synthetic and real world data. We generate synthetic trees of six different shapes. The left branch, right branch, and the zigzag tree (Figure 7) are constructed such that the strategies Zhang-L, Zhang-R, and Demaine-H are optimal, respectively; for the full binary tree both Zhang-L and Zhang-R are optimal; the mixed tree shape does not favor any of the algorithms; the random trees vary in depth and fanout (with a maximum depth of 15 and a maximum fanout of 6).

We choose three real world datasets with different characteristics. SwissProt[1] is an XML protein sequence database with 50000 medium sized and flat trees (maximum depth 4, maximum fanout 346, average size 187); TreeBank[2] is an XML representation of natural language syntax trees with 56385 small and deep trees (average depth 10.4, maximum depth 35, average size 68). TreeFam[3] stores 16138 phylogenetic trees of animal genes (average depth 14, maximum depth 158, average fanout 2, average size 95).

---

[1] http://www.expasy.ch/sprot/
[2] http://www.cis.upenn.edu/~treebank/

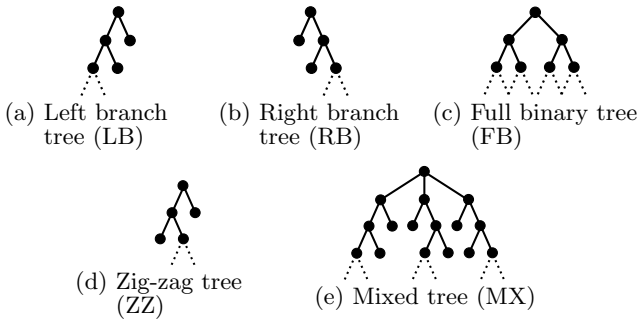

Figure 7: Shapes of the synthetic trees.

*Number of Relevant Subproblems.* We first compare the number of relevant subproblems computed by each of the algorithms for different tree shapes. The relevant subproblems are the constant time operations that make up the complexity of the algorithm. We create trees with different shapes (the five shapes shown in Figure 7 and random trees) with sizes varying between 20 and 2000 nodes. We count the number of relevant subproblems computed by each of the algorithms for pairs of identical trees. The results are shown in Figure 8. Each of the tested algorithms, except RTED, degenerates for at least one of the tree shapes, i.e., its asymptotic runtime behavior is higher than necessary. RTED either wins together with the best competitor (left branch, right branch, full binary, zigzag), or is the only winner (random, mixed), which confirms our analytic results. Note that the trees, for which RTED is even with another algorithm, were created such that the strategy of one of the competitors is optimal. The differences are substantial, for example, for the left branch trees with 1700 nodes (Figure 8(a)) Zhang-R produces 2290 times more relevant subproblems than RTED; for the mixed trees with 1600 nodes, the best competitor of RTED (Zhang-L) does 8.5 and the worst competitor (Klein-H) 30 times more computations.

*Runtime on Synthetic Data.* In Figure 9 we compare the runtimes of Zhang-L, Demaine-H, and RTED for different tree shapes. The runtime for the full binary tree is shown in Figure 9(a). Zhang-L and RTED scale well with the tree size, whereas the runtime of Demaine-H grows fast. This is expected since Demaine-H must compute many more subproblems than the other two algorithms (cf. Figure 8(c)). The strategy of Zhang-L is optimal for the full binary tree such that Zhang-L and RTED compute the same number of subproblems. The overall runtime of RTED is higher since RTED pays a small additional cost for computing the strategy. Further, our implementation of Zhang-L is optimized for the hard-coded strategy, such that the runtime per subproblem is smaller for Zhang-L than for RTED (by a constant factor below two). For other tree shapes, this advantage of Zhang-L is outweighed by the smaller number of subproblems that RTED must compute. For the zig-zag trees in Figure 9(b), Zhang-L is slower than both RTED and Demaine-H. RTED's overhead for the strategy computation is negligible compared to the overall runtime, and RTED is slightly faster than Demaine-H. For the mixed tree shapes in Figure 9(c), RTED scales very well, while the runtimes of both Zhang-L and Demaine-H grow fast with the tree size.

[3] http://www.treefam.org/

*Scalability of Similarity Join.* We generate a set of trees, $T = \{LB, RB, FB, ZZ, Random\}$, with different shapes and approximately 1000 nodes per tree. We perform a self join on $T$ that matches two trees $T_1, T_2 \in T$ if $TED(T_1, T_2) < \tau$. Table 1 shows the runtime (average of three runs) and the number of relevant subproblems computed in the join. RTED widely outperforms all other algorithms. Different from the previous experiment, where we tested on pairs of identical trees, the join computes the distance between all pairs of trees, regardless of their shape. The competitors of RTED degenerate for some pairs of trees with different shapes, leading to high runtimes. For example, both Zhang-L and Zhang-R run into their worst case for pairs of unbalanced trees LB and RB. Unbalanced trees appear frequently in practice, for example, in the phylogenetic dataset. The runtime per relevant subproblem differs between the solutions and is the smallest for Zhang-L and Zhang-R.

| Algorithm | Time [sec] | #Rel. subproblems |
|---|---|---|
| Zhang-L | 694 | 26.76 x $10^9$ |
| Zhang-R | 908 | 27.13 x $10^9$ |
| Klein-H | 2483 | 41.82 x $10^9$ |
| Demaine-H | 938 | 17.62 x $10^9$ |
| RTED | 140 | 1.96 x $10^9$ |

Table 1: Join on trees with different shapes.

*Overhead of Strategy Computation.* RTED computes the optimal strategy before the tree edit distance is computed. We measure the overhead that the strategy computation adds to the overall runtime. We run our tests on three datasets: SwissProt, TreeBank, and a synthetic data set with random trees that vary in size, fanout, and depth. We pick tree pairs at regular size intervals and compute the tree edit distance. For a given tree size $n$ we pick the two trees in the dataset that are closest to $n$; the size value used in the graphs is the average size of the two trees. Figure 10 shows the runtime for computing the strategy for a pair of trees and the overall runtime of RTED. The strategy computation scales well and the fraction it takes in the overall runtime decreases with the tree size. The spikes in the overall runtime are due to the different tree shapes, for which more or less efficient strategies exist. The runtime for the strategy computation is independent of the tree shape. The results confirm our analytic findings in Section 6.

*Scalability on Real World Data.* We measure the scalability of the tree edit distance algorithms for the TreeFam dataset with phylogenetic trees. We partition the dataset by size (less than 500, 500–1000, more than 1000 nodes) and compute the distance between pairs of trees of two partitions. We measure the performance of RTED as the percentage of relevant subtree computations that RTED performs with respect to the best (Table 2(a)) and the worst competitor (Table 2(b)). The best and worst competitors vary between the pairs of partitions. The tables show the results for random samples of size 20 from each partition. RTED always computes less subproblems than all its competitors (84.2% to 94.4% w.r.t. the best competitor, 5.6% to 30.6% w.r.t. the worst competitor). The advantage increases with the tree size. For the partitions with the largest trees, RTED

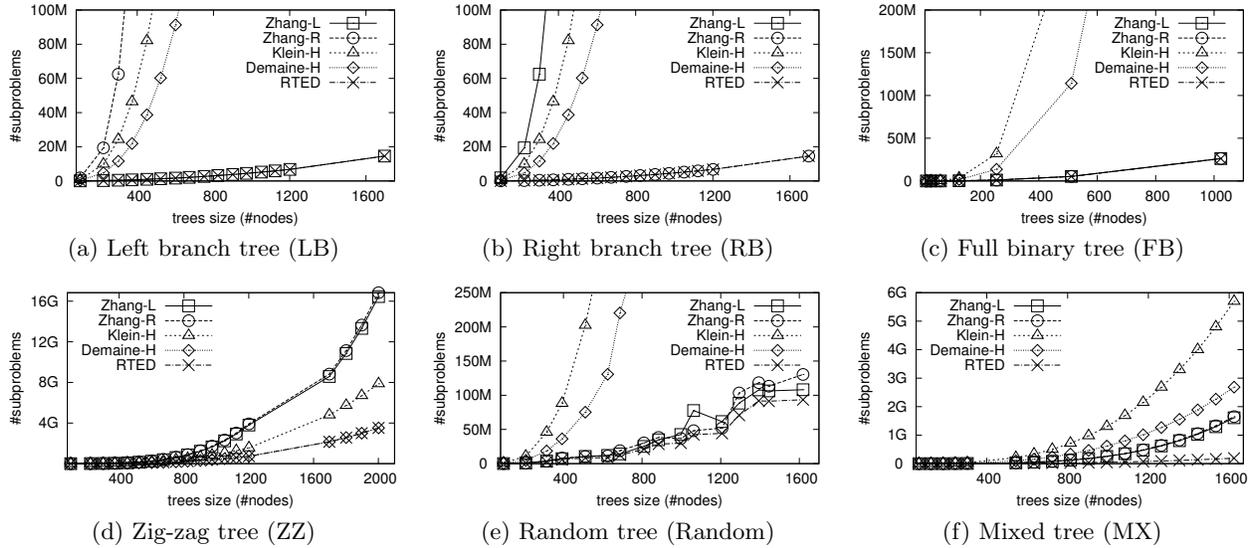

Figure 8: Number of relevant subproblems for different algorithms and tree shapes.

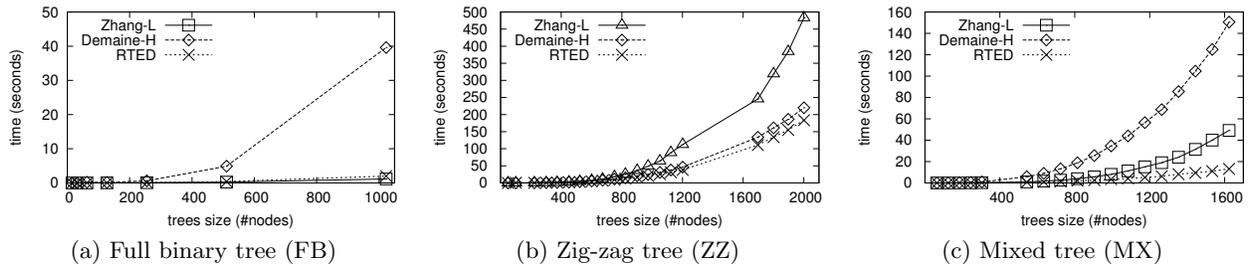

Figure 9: Runtime of the fastest tree edit distance algorithms for different tree shapes.

(a) RTED to the best competitor

| ↓ tree sizes → | <500 | 500-1000 | >1000 |
| --- | --- | --- | --- |
| <500 | 94.4% | 90.1% | 88.2% |
| 500-1000 | 91.4% | 85.6% | 84.2% |
| >1000 | 89.4% | 85.9% | 86.9% |

(b) RTED to the worst competitor

| ↓ tree sizes → | <500 | 500-1000 | >1000 |
| --- | --- | --- | --- |
| <500 | 30.6% | 17.6% | 17.3% |
| 500-1000 | 21.1% | 8.9% | 7.9% |
| >1000 | 18.3% | 7.7% | 5.6% |

Table 2: Ratio of relevant subproblems computed by RTED w.r.t. the (a) best and (b) worst competitor.

produces 18 times less relevant subproblems than the worst competitor. This experiment shows the relevance of RTED in practical settings. The wrong choice among the competing algorithms for a specific dataset may lead to highly varying runtimes. RTED is robust to different tree shapes and always performs well.

## 9. CONCLUSION

In this paper we discussed the tree edit distance between ordered labeled trees. We introduced the class of LRH strategies, which generalizes previous approaches and includes the best algorithms for the tree edit distance. Our general tree edit distance algorithm, GTED, runs any LRH strategy in $O(n^2)$ space. We developed an efficient algorithm for computing the optimal LRH strategy for GTED. The resulting algorithm, RTED, runs in $O(n^2)$ space as its best competitors and the $O(n^3)$ runtime of RTED is worst-case optimal. Compared to previous algorithms, RTED is efficient for any tree shape. In particular we showed that the number of subproblems computed by RTED is at most the number of subproblems that its best competitor must compute. Our empirical evaluation confirmed that RTED is efficient for any input and outperforms the other approaches, especially when the tree shapes within the dataset vary.

**Acknowledgements.** Part of the work has been done in the context of the RARE project, which is funded by the Autonomous Province of Bolzano - South Tyrol, Italy. We thank Periklis Andritsos for his valuable comments.

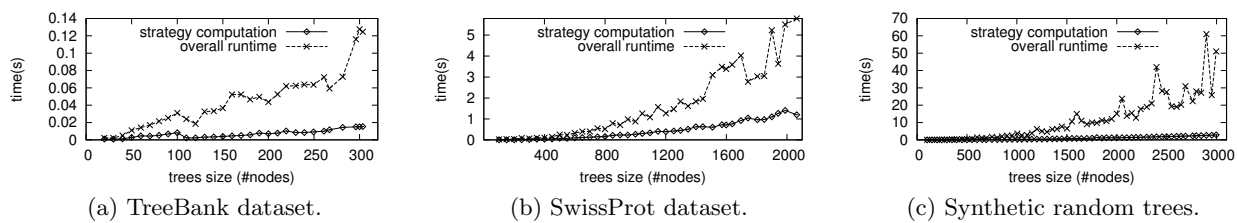

Figure 10: Overhead of the strategy computation in the overall execution time of RTED.